\documentclass{article}

\usepackage[preprint]{neurips_2024}

\usepackage[utf8]{inputenc} 
\usepackage[T1]{fontenc}    
\usepackage{hyperref}       
\usepackage{url}            
\usepackage{booktabs}       
\usepackage{amsfonts}       
\usepackage{nicefrac}       
\usepackage{microtype}      
\usepackage{xcolor}         

\title{Unsettled Law: Time to Generate New Approaches?}

\author{
    David Atkinson \\
    Assistant Professor of Instruction \\
    Business, Government and Society Department \\
    McCombs School of Business \\
    University of Texas, Austin \AND
    Jacob Morrison \\
    Allen Institute for AI
\\ \AND
 \small{
   \textbf{Correspondence:} \textnormal{\{davida, jacobm\}@allenai.org}
 }
}

\begin{document}

\maketitle

\begin{abstract}
    We identify several important and unsettled legal questions with profound ethical and societal implications arising from generative artificial intelligence (GenAI), focusing on its distinguishable characteristics from traditional software and earlier AI models. Our key contribution is formally identifying the issues that are unique to GenAI so scholars, practitioners, and others can conduct more useful investigations and discussions. While established legal frameworks, many originating from the pre-digital era, are currently employed in GenAI litigation, we question their adequacy. We argue that GenAI's unique attributes, including its general-purpose nature, reliance on massive datasets, and potential for both pervasive societal benefits and harms, necessitate a re-evaluation of existing legal paradigms. We explore potential areas for legal and regulatory adaptation, highlighting key issues around copyright, privacy, torts, contract law, criminal law, property law, and the First Amendment. Through an exploration of these multifaceted legal challenges, we aim to stimulate discourse and policy considerations surrounding GenAI, emphasizing a proactive approach to legal and ethical frameworks. While we refrain from advocating specific legal changes, we underscore the need for policymakers to carefully consider the issues raised. We conclude by summarizing key questions across these areas of law in a helpful table for easy reference.
\end{abstract}

\section{Introduction}

\subsection{The Law of the Horse}

In 1996, Judge Frank H. Easterbrook gave a presentation at a conference called “Law of Cyberspace” on what he called the law of the horse and he subsequently published his argument \cite{easterbrook1996}. In it, he noted that courses with titles like “Law and \_\_\_” were unnecessary\footnote{Such focus was for “dilettantes”} and that any coupling of law with something else should reveal something meaningful about law as a whole and not just a niche area. 

Judge Easterbrook also predicted that many prognostications about new technologies would not materialize, so efforts spent focusing on how to conform the law to them would be in vain. There was no need for “Torts and Cyberspace” or “Crime and Cyberspace.” Rather than think about how cyberspace may be different, he argued, one should focus on improving and clarifying existing laws and then simply apply them to cyberspace the same way one might apply them to cars, or houses, or horses. Moreover, the law around technologies already in existence in 1996 was still unsettled. If those laws and their attendant rights were murky, what hope or use was there in further morphing the law for technology that was in its infancy?

\subsection{What Happened}

With hindsight, we know that in the 28 years since Judge Easterbrook gave his speech we have continued to apply traditional tort, intellectual property, and contract laws, but we have also incorporated some novel laws of cyberspace. This includes the DMCA (Digital Millennium Copyright Act), CFAA (Computer Fraud and Abuse Act), COPPA (Children’s Online Privacy Protection Act), and ECPA (Electronic Communications Privacy Act). The internet, clearly, was a difference in kind from physical reality in some ways, not merely a difference in degree. 

In 1999 Professor Larry Lessig wrote \cite{lessig1999} the most famous rebuttal to Judge Easterbrook’s stance, challenging the notion that cyberspace was not meaningfully different from the physical world, and Lessig again summarized this view in 2021 \cite{lessig2021}: 

\begin{quote}
    “I’m a believer in the value of settled doctrine. I agree that at the margin, the question is always and only how existing doctrine applies to a new set of facts. Fidelity to role requires that courts minimize interpretive depth. Rules must be clear and applied in obvious ways if the very act of applying rules is not to be read as inherently political. \textit{But there are exceptions to this conservative principle, at least when technology has changed fundamentally, and when the consequences of that change for important institutions or traditions are severe.}” (emphasis added)
\end{quote}

To see how the law typically adapts to new technology, consider how Judge Easterbrook himself led to a watershed moment in contract law with ProCD, Inc. v. Zeidenberg, 86 F.3d 1447 (7th Cir., 1996), which held that terms were enforceable even if a buyer could not see them until after buying and removing the shrinkwrap from a CD-ROM. This was a stark change from how contract law began in the US, where a contract was only enforceable if the parties entered a voluntary agreement through a bargained-for exchange. Then, in the mid-1900s, contract law shifted to accommodate contracts of adhesion for mass produced goods where people still had to sign an agreement to make a purchase, showing that they had time to review the terms. But Judge Easterbrook, writing for the majority, broadened the aperture further, binding people to agreements they could not even read prior to purchasing the goods, and the terms automatically applied the moment they started using the goods. From this came the application to the internet with what is known today as “browsewrap” (when the terms, typically linked in the footer, automatically apply to those who use a website) and “clickwrap” (which requires people to affirmatively click their acceptance of the terms in order to proceed or access a website), which will be further discussed below.

The areas regulators chose to not treat differently from cyberspace are also telling, as it led to a lack of regulation around social media, facial recognition, the collection of personal data, and more, for example, which has contributed to fundamentally new forms of economic incentives in areas like privacy. We also see how incentives from regulation (or a lack thereof) have affected contract law, as noted above, online protections for children, and copyright law, for example. 

The question is whether the law is once again stretching past a reasonable interpretation of how it was written in restatements from 40+ years ago\footnote{Restatement (Second) of Contracts has a copyright of 1981} and whether courts and regulators should adapt accordingly. We will explore what the technology looked like when the relevant laws were passed, and then we’ll discuss whether it makes sense to have something like a law of GenAI to supplement or replace some of the last twenty years’ law of cyberspace.

\section{Technical Differences}

This section will explore three primary areas of software capabilities and advancement to demonstrate its progression. 

\subsection{Traditional Software}

It’s helpful to think of this as “Good Old Fashioned AI,” or GOFAI. Broadly, one can think of this type of AI being the state of the art and most fashionable form from the 1950s to 2012.

Traditional software operated on deterministic principles, meaning that given a specific set of inputs, the software will produce the same outputs every time. This predictability is achieved through well-defined algorithms and rules that govern the software's behavior that is typically hard-coded by developers. The software's behavior is entirely defined by the code written by programmers, which specifies how to handle various inputs and scenarios.

The computational requirements of traditional software are generally modest, particularly when compared to modern applications involving artificial intelligence or big data. Traditional software programs often execute in environments with limited processing power and memory, such as desktop applications or embedded systems, (hopefully) encouraging developers to optimize their code to be efficient and resource-conscious.

The complexity and scale of traditional software is often measured in terms of the number of lines of code (LOC), with the largest codebases in the world ranging from tens of millions to billions of LOC\footnote{\href{https://www.wired.com/2015/09/google-2-billion-lines-codeand-one-place/}{https://www.wired.com/2015/09/google-2-billion-lines-codeand-one-place/}}. This metric provides a rough estimate of the software's size and complexity, with larger codebases generally indicating more complex or comprehensive functionality. However, a higher LOC count doesn't always mean better or more advanced software, as it can also indicate redundancy or lack of optimization.

\subsection{“Traditional” AI}

This is the form of AI that garnered so many headlines in the 2010s with advances in facial recognition and recommendation algorithms. Broadly, one can think of this type of AI being the state of the art and most fashionable form from 2012ish to 2022ish.

Before the advent of GenAI, artificial intelligence systems were largely designed to recognize and respond to specific patterns within data, using statistical methods like deep learning, to make decisions or predictions. Unlike traditional software, which is deterministic, these AI systems were probabilistic, meaning their outputs were based on likelihoods rather than certainties.

The models used in pre-GenAI were relatively small compared to today's standards, containing thousands to hundreds of millions of parameters (the adjustable weights within the model that are tuned during training). They were typically designed for narrower applications, such as task-specific image recognition, spam detection, and recommendation systems. These models focused on specific tasks within defined domains, relying on specialized algorithms to process input data and produce outputs. 

The learning process was predominantly supervised, meaning the models were trained on datasets where the correct outputs were already known. Thus training these models required a substantial amount of labeled data, often ranging from tens of thousands to millions of examples. This data was essential for the models to learn the patterns and relationships within the input data, and the labeling process was generally slow and expensive, or relied on noisy signals within the data itself (like emojis in tweets for sentiment analysis).

\subsection{Generative AI}

This is the form of AI that has captured the public’s imagination since ChatGPT launched in late 2022. Broadly, one can think of this type of AI being the state of the art and most fashionable form today.

GenAI represents a significant advancement in artificial intelligence, characterized by its ability to create new content, such as text, images, and video, that mimics human-like creativity. Similar to traditional AI models, GenAI operates probabilistically and generally functions as a "black box," meaning that the internal workings and decision-making processes are not easily interpretable by humans. While there is much work being conducted to interpret the inner workings of the largest models today\footnote{\href{https://www.anthropic.com/news/golden-gate-claude}{https://www.anthropic.com/news/golden-gate-claude}}, it is a comparatively less developed area of research.

Top-tier GenAI models, such as Claude, Gemini, and GPT-4, are massive in scale, with billions to trillions of parameters, and they generally undergo a complex, multi-staged training process \cite{geminiteam2024gemini, openai2024gpt4}. The immense size of these models allows them to capture a vast amount of information and nuanced patterns within the data, enabling them to generate highly coherent and contextually relevant outputs.

Training generative AI models requires an enormous amount of data, often measured in trillions of tokens (pieces of text, patches of images, video frames, etc). This data comes from diverse and extensive sources, allowing the models to “learn” a wide range of language patterns and concepts. Unlike traditional supervised learning, GenAI models typically use self-supervised learning, a method where the model learns to predict parts of the data from other parts. For example, in language models, the model learns to predict the next word in a document based on the preceding words, allowing it to learn contextual relationships without needing explicitly labeled datasets.

The computational demands for training generative AI models are significantly higher than those for earlier AI systems. The training process involves extensive parallel computations across multiple GPUs or specialized hardware, requiring vast amounts of computational power and energy. This increased need for compute resources is due to the large model size and the complexity of the training algorithms, and is of growing concern in the public eye\footnote{\href{https://www.forbes.com/sites/cindygordon/2024/02/25/ai-is-accelerating-the-loss-of-our-scarcest-natural-resource-water/}{https://www.forbes.com/sites/cindygordon/2024/02/25/ai-is-accelerating-the-loss-of-our-scarcest-natural-resource-water/}}.


\section{New Capabilities From New Technology}

Our focus is on why certain areas of the law may need to be modified in significant ways; it’s not intended to identify all the possible implications of GenAI. However, it may be helpful to quickly identify some significant capabilities that GenAI enables that previous types of AI and traditional software programs could not do at scale, as quickly, as easily, as cheaply, or perhaps at all just a few years ago:

\begin{itemize}
    \item Creating realistic-looking deepfake images and voices.
    \item Replicating creator styles (artistic, writing)
    \item Creating misinformation and disinformation
    \item Adapting to user conversations in a way that seems human
\end{itemize}

Each of these capabilities has implications for multiple areas of law, and they each require a rethinking of not only how to deal with outputs from the model, but also how the data it’s trained on is collected, how it’s trained, and how the model is deployed.

\section{Should the Law Treat GenAI Differently?}

While the technical differences between traditional software, neural nets over the past decade, and GenAI from the past few years may be profound, that does not necessarily imply the law should treat them differently. 

Though many foundational laws were passed a half century ago or more, or they rely on case law that initially developed centuries ago, they continue to be relevant today. This is evidenced by how litigants are seemingly finding no shortage of claims to bring against GenAI companies \cite{atkinson2024legalrisktaxonomygenerative}. Rather than advocate for new laws and regulations, maybe we should continue applying the tools we currently have and let the courts fine-tune how laws are applied. This has many advantages, including predictability, stability, and familiarity. 

On the other hand, just as we needed to create new laws when cars came into widespread use and replaced the horse and buggy to control speeding, intersections, inspections, fueling, safety, and more, and that led to the reshaping of state and federal budgets, the arrangements of cities, and so on, it may be appropriate to give our legal system a significant renovation to properly address the genuine novelty of GenAI\footnote{The car is just one example. Here’s another: we didn’t want a right to be forgotten until there was cheap, reliable storage. New technology introduced new issues that required novel legal and regulatory solutions.}. Such novelty arises from GenAI’s unique ability–relative to other forms of software–to be general purpose, its reliance on several orders of magnitude more data and compute, its ability to perform several tasks as well as humans, its inscrutable internal workings, and the pace of improvement and application in society, to name a few fundamental and likely highly-consequential advances. 

For example, in some instances, as with copyright, it may be that the current law is insufficient. A company can follow the letter of the law while plainly ignoring the spirit of it, undermining its purpose and leaving little or no recourse for those who most of society feel were wronged. Yet companies can hardly be blamed for pressing the boundaries of what they can legally do. That's sometimes the area where innovation lives and is surely where high market valuations can be found. 

While we do not take a firm stand on what changes, if any, need to occur for any particular body of law, we do strongly believe policymakers should give the following topics due consideration now and err on the side of being proactive rather than reactive, as recent history around privacy and antitrust has shown that hesitating to take action often leads to never taking action. What follows are some representative examples of unsettled legal issues that may need to be significantly modified for GenAI in the near term, rather than wait for incremental changes.

\subsection{Copyright Law}

When thinking about whether we need new law for GenAI, perhaps we should consider not what copyright is today and whether it applies as written, but whether we've stretched copyright law too far, first from the media of the 1970s to cyberspace, then from cyberspace to AI, and now to possibly absurd lengths for GenAI. Rather than focus on the written law, we could instead focus on the law’s purpose, which is to further Article I Section 8 of the Constitution: “To promote the Progress of Science and useful Arts, by securing for limited Times to Authors and Inventors the exclusive Right to their respective Writings and Discoveries.” 

Of course, GenAI may be an overall force for good. It may enable more people to partake in creative enterprises, marketing, advertising, idea iteration, and so on. Previously, larger companies with more money and people could do all of the above faster and in greater volume, but perhaps GenAI levels the playing field, giving people and entities with fewer resources significantly greater abilities, thereby helping to promote science and art.

Assuming copies of copyrighted material is infringement, it may be that GenAI outputs are transformative compared to the training data. It may also be that the outputs do not directly compete with the copyright owners. However, it may also be the case that the outputs undermine progress of science and useful arts by obviating the need for users to ever visit the original source material, thereby depriving the content creators of ad revenues, the ability to sell products, the ability to build a brand or reputation, or the ability to entice GenAI users to sign up for a subscription. Even when potential purchasers visit sites, it’s cold comfort when human works are obscured by the voluminous outputs of bots, such as on Etsy\footnote{\href{https://www.theatlantic.com/technology/archive/2023/06/ai-chatgpt-side-hustle/674415/}{https://www.theatlantic.com/technology/archive/2023/06/ai-chatgpt-side-hustle/674415/}}\footnote{\href{https://www.pcmag.com/news/is-ai-ruining-etsy-loosening-definition-of-handmade-frustrates-artists}{https://www.pcmag.com/news/is-ai-ruining-etsy-loosening-definition-of-handmade-frustrates-artists}}\footnote{\href{https://www.nbcnews.com/tech/tech-news/etsy-crochet-buyers-suspect-ai-made-images-used-sell-patterns-rcna145878}{https://www.nbcnews.com/tech/tech-news/etsy-crochet-buyers-suspect-ai-made-images-used-sell-patterns-rcna145878}}. Without sufficient incentives, and with insufficient traffic visiting their sites, content creators may be discouraged from either creating or sharing new content, or they may not have the revenue necessary to sustain content creation operations even if they’d like to continue creating and sharing. If so, this would suggest a radical rethinking of copyright law as written may be in order. Notably, no form of software has ever had the capacity to wither entire sectors of the economy the way GenAI could, and such changes may be largely irreversible.

\subsubsection{Fair Use Factors}

Copyright law may need more than a light touch-up even if we decide to keep most of it in place as is. GenAI leans heavily on the idea of fair use, for example, in a way that no other technology has. Unlike VCRs\footnote{See Sony Corp. of Am. v. Universal City Studios, Inc., 464 U.S. 417 (1984)}, large GenAI models must infringe to exist because GenAI companies do not have authorization to use most copyrighted works they copy, the developers must copy all the copyrighted work they collect for training datasets, and then they must copy those copies into the model to train the model. Absent authorization, the copying of copyrighted data infringes on the copyright owner’s exclusive right to reproduce their work. Therefore, it seems the only way around this at a scale that is workable is the affirmative defense of fair use. 

Of the four fair use factors (the purpose and character of the use, the nature of the copyrighted work, the amount or substantiality of the portion used, and the effect of the use on the potential market for or value of the work), factors two and three are even less relevant to GenAI than is typical in copyright defenses generally, and typically those two factors are the least important in a fair use analysis\footnote{See, e.g., Bill Graham Archives v. Dorling Kindersley Ltd., 448 F.3d 605 (2d Cir. 2006) “the second factor may be of limited usefulness where the creative work of art is being used for a transformative purpose.”; Bell v. Eagle Mt. Saginaw Independent School District U.S. Court of Appeals, Fifth Circuit, 27 F.4th 313 (2022) “The nature of the work is widely considered the least significant fair-use factor.”}. The remaining factors will apply similarly in almost all uses of GenAI, so a new law that specifically addresses GenAI may be far more efficient than allowing different districts in different states to litigate substantially similar fact patterns repeatedly and likely occasionally arrive at different conclusions, inefficiently keeping the market unpredictable. 

For example, some GenAI products, like Google’s AI Overview and Perplexity.AI’s summarization engine are based on visiting mostly copyrighted content and summarizing it. Such automated summarizing was not possible on an internet-wide scale prior to GenAI. A refined version of fair use’s applicability could make analyses of claims like misappropriation of hot news, where a competitor summarizes a publisher’s material before the publisher has an opportunity to commercially benefit from it, or if the publishing undermines the point of subscribing to the publisher, more efficient and make it clearer when such summarizing is clearly legal.

\subsubsection{People vs. GenAI}

Any allowances for infringement by companies due to fair use may mean GenAI is given greater rights than people. GenAI’s most capable models require billions of files, which likely include essays, short stories, novels, poems, lyrics, and more. One infamous training dataset, Book3, contains nearly 200,000 books reproduced without authorization\footnote{\href{https://www.theatlantic.com/technology/archive/2023/09/books3-database-generative-ai-training-copyright-infringement/675363/}{https://www.theatlantic.com/technology/archive/2023/09/books3-database-generative-ai-training-copyright-infringement/675363/}}. Other sites likely scraped for GenAI training material, like LibGen, contain over 4.5 million books and tens of millions of other copyrighted materials reproduced and distributed without authorization\footnote{\href{https://libgen.onl/}{https://libgen.onl/}}. In contrast, if a human made an unauthorized copy of every book they read in their lifetime to learn, few people would copy more than a few thousand.This means fair use for AI companies is expanding and applying well beyond anything a human could possibly claim similar protections for.  

Some scraping has been common for decades. However, in those instances, such as with Google search, the benefit to society or content creators was arguably much clearer. When Google indexed sites, it was scraping to help other people find those sites, which was mutually beneficial to both Google and the sites. When Google copied books, it was to help people discover facts about the books, Google only revealed short snippets from the books, and the search results linked people to the full text on authorized sites. For other scraping uses, such as for researchers, the volume of content scraped was historically much less, it was used for non-commercial purposes, and it was generally only used for a specific, narrow purpose \footnote{The SQuAD dataset \cite{rajpurkar2016squad100000questionsmachine, rajpurkar2018knowdontknowunanswerable} is a good example. It is still popular, but it only has around 150,000 examples of scraped materials.}. 

If the argument is that GenAI’s use of the copyrighted material is transformative and non-competitive, the same could apply to humans. In fact, it may apply even more to humans. GenAI has not yet developed a new, profound scientific theory, for example, but humans have and do. Humans would probably develop even more such theories if anyone was able to access and read any copyrighted work they wanted without fearing liability for copyright infringement. 

Humans are also far less likely to memorize large chunks of text like GenAI and are probably no more likely to paraphrase or plagiarize, and they certainly won’t produce it at the same scale\footnote{\href{https://copyleaks.com/about-us/media/copyleaks-research-finds-nearly-60-of-gpt-3-5-outputs-contained-some-form-of-plagiarized-content}{https://copyleaks.com/about-us/media/copyleaks-research-finds-nearly-60-of-gpt-3-5-outputs-contained-some-form-of-plagiarized-content}}. In short, it’s not clear why any allowance for GenAI to access content on the basis that its output will usually be transformative and beneficial should not also extend to humans. In sum, if the law overextends in a way that grants GenAI more legal protections than humans, it seems it should be accompanied by a strong justification.

Such an extension may do more harm than good, though. If any person and any machine can access any copyrighted material without having to pay, for instance, it would likely cause a collapse in several industries as wide-ranging as music to textbook publishing. A discussion around fair use exemptions should investigate whether disparities in data access and creative output between humans and GenAI justify differing legal standards.

\subsubsection{Data Provenance}

There is also great uncertainty around whether the provenance of the data matters. If developers know or should know that their training dataset includes data, such as books, that were taken without the copyright owner’s authorization, does that matter? Or does fair use absolve every controversial source of data? Does it matter if the developers know or should know about the data’s sketchy origins, or should the standard be strict liability, as it typically is with copyright law? Should data intended to be publicly available be treated the same as data that is publicly available but wasn’t intended to be by the data creator?

If the provenance does not matter so long as the AI company itself did not create the initial dataset (ie., did not circumvent technological measures to prevent unauthorized access, or “hack” into a website), might that incentivize even more unauthorized taking of copyrighted material that could be laundered into AI training datasets?

\subsubsection{A New Secondary Liability?}

To the extent that a court may find that GenAI developers did not directly infringe on a copyright when a user coaxes a substantially similar output from the model, a court may look at whether the GenAI entity committed secondary liability. It may be that under the established doctrines there would be no secondary liability for any number of highly plausible reasons\footnote{For example, it’s not clear how it would apply to open-source models, where there is no financial benefit and the open-sourcing entity can no longer control the model or what people can do with it.}, but that result may be unsatisfactory for many. 

Copyright law may need a revamped version of secondary liability analysis that incorporates a negligence regime, as UT law professor Oren Bracha has proposed\footnote{Oren Bracha, Generative AI’s Two Information Goods (forthcoming)}, so that if there were reasonable precautions that were not unreasonably burdensome to implement to prevent any infringement, but developers didn’t implement them, then the GenAI entity could be found liable for copyright infringement even if they would not be liable under existing secondary liability schemes. 

\subsubsection{How Strict?}

Determining the threshold for copyright infringement in GenAI outputs demands nuanced analysis. Factors as varied as output similarity, frequency of infringement, and prompting techniques employed to elicit infringing outputs could each individually affect a determination of infringement.

It may be that a model can produce substantially similar outputs to the material it was trained on (i.e., the copyrighted work), but even so, does the frequency of infringing outputs matter, or is a single output sufficient for infringement? For example, if a user must prompt a model one million times to get a single infringing output should that be viewed differently from a model that produces an infringing output every ten prompts? Does it matter if the infringing outputs are only possible if the user applies a jailbreaking technique?

\subsection{Privacy}

As of March 18, 2024, only 15 states had comprehensive data privacy laws\footnote{\href{https://pro.bloomberglaw.com/insights/privacy/state-privacy-legislation-tracker/}{https://pro.bloomberglaw.com/insights/privacy/state-privacy-legislation-tracker/}}. Instead of a complex patchwork stretching from coast to coast, citizens and companies might be better served by a federal privacy law. A federal law would allow everyone to understand their rights, obligations, and limitations. One’s health information is no less valuable to data brokers or developers making a medical-focused large language model just because they live a few feet on one side of a state border versus another, and it probably does not make sense to treat such sensitive matters as if it’s just a matter of preference or ideology. 

\subsubsection{Linking Data}

While privacy concerns around data aren’t new, they take on a different flavor with huge datasets. In the 1800s privacy was mostly focused on what happened in a person's house and family affairs \cite{gajda2022}. But by 2016 the Supreme Court declared that “Indeed, a cell phone search would typically expose to the government far more than the most exhaustive search of a house: A phone not only contains in digital form many sensitive records previously found in the home; it also contains a broad array of private information never found in a home in any form.”\footnote{Riley v. California, 573 U.S. 373 (2014)}

Given that GenAI requires the vacuuming up of orders of magnitude more data than was necessary for previous types of artificial intelligence, it may resemble a globe-spanning smartphone more than an Excel sheet or image classifier dataset. And given its broad nature, it’s unsurprising the datasets contain personal data, including sensitive information such as health, financial, sexual orientation, religion, and other forms of data. Collecting data from the internet writ large means it may be easier to link data together, revealing more than could be discovered by coming across each type of information in isolation when it’s more obscured. It’s not clear that when people post on the internet (and especially when they made the post several years ago) that they assume “viewable to the public” also meant “free to collect and process by AI companies.”

Of note, people have always assumed no other private entities can search a person whenever they want, looking for whatever they want. Part of the reason the Fourth Amendment of the Constitution was written was to make it clear that even the government cannot search citizens with general writs of assistance\footnote{\href{https://constitution.congress.gov/browse/essay/amdt4-2/ALDE\_00013706/}{https://constitution.congress.gov/browse/essay/amdt4-2/ALDE\_00013706/}}, but instead must acquire a warrant based on “probable cause, supported by Oath or affirmation, and particularly describing the place to be searched and the persons or things to be seized.”\footnote{United States Constitution amendment IV} Allowing anyone to scrape the entire public internet for the sake of AI, including, presumably, the government, may start to fall into Riley v. California smartphone territory.

\subsubsection{Unknown Content}

Currently, none of the AI entities that develop the most powerful GenAI models have revealed the contents of their datasets. Unless you happen to be one of the handful of people in the world granted access to the entities’ training data, there is no way to establish with certainty how many times a piece of personal data is in the training data, from where the personal data was collected, how many pieces of personal data the AI entity has about any particular individual, or what steps were taken to not collect or to remove such personal data from the training datasets. 

We may want to explore mechanisms for enhancing transparency and accountability. This could include examining the feasibility of regulatory scrutiny, independent audits, and public disclosure requirements to ensure the responsible and ethical handling of personal data. But this goal should be balanced against the privacy risks of transparency: if anyone can poke around in a dataset that includes personal data, might that introduce more risks than an opaque dataset with limited access?

\subsubsection{Inability to Comply with GDPR}

This section will use the General Data Protection Regulation (GDPR) as a representative of privacy laws. GDPR grants people several privacy rights\footnote{\href{https://gdpr-info.eu/}{https://gdpr-info.eu/}}, including the rights of access, rectification, erasure, restriction of processing, data portability, and to object, among others. There are some carve outs and exceptions, but they generally don’t apply to commercial enterprises that collect and process data without notice or consent. 

The issue is that GenAI entities usually cannot comply with these types of privacy rights even if they wanted to. The technical limitations and complexity of GenAI models means it’s extremely difficult–if not impossible–to identify where any individual piece of personal data resides within a model’s weights. This means, for example, that the entities cannot simply delete the data from the model. Instead, they rely on other techniques to obfuscate the data, rather than remove it, such as RLHF and applying filters. The same applies to other rights. If the model trained on incorrect data or outputs incorrect data about an individual as if it’s fact, there is no practical way to directly rectify it within the trained model.

When an AI entity says they cannot comply with the law, but can provide an outcome that mostly meets the spirit of the law, is it sufficient, or is that just another way of an AI entity saying they get to choose which laws they must comply with and in which fashion? Should regulators prohibit the deployment of technology that cannot comply with laws as written (taking a firm stance on what is a “right,” such as a right to privacy, versus what is a “privilege”), or should the laws adapt to technical realities?

\subsection{Torts}

When traditional software or traditional AI goes wrong, it goes wrong for a specific task. A bug in a calculator app may cause the program to think 2x2 = 5. More powerful tools could only create one type of powerful harm. For example, facial recognition software could be misused to inappropriately recognize faces, but it won’t tell you 2x2=5. Or 4, for that matter. But GenAI has far greater breadth as a general purpose tool and can therefore cause far more harms.

Currently, torts does not always cover a vast swath of harms that were difficult or impossible to create prior to GenAI. For example, the scourge of non consensual deepfake porn, especially when targeting minors, is largely unaddressed by existing common and statutory law. Only a handful of states have a law preventing such AI generations based on GenAI\footnote{\href{https://www.nytimes.com/2024/04/08/technology/deepfake-ai-nudes-westfield-high-school.html}{https://www.nytimes.com/2024/04/08/technology/deepfake-ai-nudes-westfield-high-school.html}}. These types of realistic images were impossible prior to the massive data collection and computing power of recent years, so perhaps such laws were unnecessary. But today it is difficult to argue a few states passing some laws that provide varying levels of protection for victims is sufficient.

As with privacy, there may be a need for a federal right to publicity. While a patchwork of case law and statutes may have sufficed when technology was more limited and less accessible, the absence of some protections on a national scale is proving to be a weakness. The ease with which someone can replicate a person's looks, voice, artistic style, and more are significantly easier and far more realistic today due to advances in GenAI than ever before and some may argue that it’s negligence that allows such alleged harms to proliferate\footnote{See, e.g. \href{https://apnews.com/article/new-hampshire-primary-biden-ai-deepfake-robocall-f3469ceb6dd613079092287994663db5}{https://apnews.com/article/new-hampshire-primary-biden-ai-deepfake-robocall-f3469ceb6dd613079092287994663db5}}.

\subsubsection{Liability Formula}

The harms of AI are still best understood under the traditional Learned Hand formula as applied to the traditional elements of tort law. The Learned Hand formula states that a party is liable if the burden of precautions is less than the product of the likelihood of harm and the magnitude of the harm. That is, if the likelihood of an AI model harming someone times the magnitude the harm if it occurs is less than the burden of preventing the harm, then the AI developer would be liable.

Because torts can apply to a wide range of harms and GenAI can potentially cause a significantly greater breadth of harms than virtually any other software, the question of prevention and mitigation of foreseeable harms is especially important. Also, it is possible that the more powerful the AI becomes in each domain (vision, text inputs and outputs, audio inputs and outputs, video inputs and outputs, etc.) the greater the magnitude of potential harm. 

However, it’s not clear what the reasonable measures of preventing harm are precisely because general purpose technologies can perform so many purposes. Courts often look at what's customary in a field, but because GenAI is relatively new, and the largest AI entities are often cagey about any preventative measures they implemented, it’s not clear what’s customary. Most GenAI processes are treated as proprietary and trade secret, so it’s not clear what entities mean when they say they “clean” their data or apply filters or conduct red teaming. And if a tort case goes to trial, it will be exceedingly difficult to know whether the model was developed or deployed negligently because the development and deployment process is so technically complex and could involve multiple actors at every stage of development.

Should torts law adjust to treat GenAI with strict liability (discussed further below), to account for the greater potential for unknowable risk to the user? Or should the standard shift from “foreseeable” to “known or knowable,” setting a higher bar for liability for GenAI developers who can’t possibly foresee all potential harms? Should companies have to follow a precautionary principle? Should a set of minimum standards be set for what’s reasonable and failing to satisfy that criteria shifts the standard from negligence to strict liability?

\subsubsection{Product Design Liability}

As traditionally applied, product liability tends to only play a role when there is physical harm\footnote{Winter v. G.P. Putnam's Sons, 938 F.2d 1033 (9th Cir. 1991)}. Courts have resisted considering information to be a product for which products liability could apply\footnote{See, e.g., Winter, 938 F.2d at 1034; Herceg v. Hustler Mag., Inc., 565 F. Supp. 802, 803 (S.D. Tex. 1983); see also RESTATEMENT (THIRD) OF TORTS: PRODUCTS LIABILITY § 19, cmt d (AM. L. INST. 1965) (distinguishing between a book as a physical product and harmful information conveyed by a book).}. In the case of books, judges don’t ask if the information contained in the books was accurate. Rather, they ask whether the books were readable (their intended purpose), and find that they are indeed legible, so the books meet their intent\footnote{Walter v. Bauer, 439 N.Y.S.2d 821, 823 (N.Y. Sup.Ct. 1981).}. Similarly, with vehicles the manufacturer can know why the vehicle does or does not do something. 

This is not the case with GenAI outputs. Users of the models usually don’t have a way to understand why a model creates any particular output and the user usually can’t modify the model even if they were determined to divine the inner workings of the machine. If anyone is in a position to sufficiently mitigate or prevent harms from foreseeable uses, it’s probably the developers. To the extent they can’t, should they not be allowed to deploy the model?

Moreover, the concept of information being largely immune from product liability may make less sense with GenAI models. A book is a one-way communication, but a model is interactive. This interactivity could lead the model to feeding users information it thinks the users want to receive\footnote{Models have no conception of right or wrong, accurate vs. merely plausible. They are instead generally trained to be helpful and not harmful. See, e.g., \href{https://www.bbc.com/news/technology-67012224}{https://www.bbc.com/news/technology-67012224}. The \textit{New York Times} has also recently picked up on this idea: \href{https://www.nytimes.com/2024/05/14/technology/ai-chatgpt-her-movie.html?smid=nytcore-android-share}{https://www.nytimes.com/2024/05/14/technology/ai-chatgpt-her-movie.html?smid=nytcore-android-share}}. A book is written with a broad audience in mind, but GenAI can act on an individual basis, relying on past interactions and information provided by individuals to personalize outputs. That is, in fact, one of its primary selling points for some frontier models. Does the interactivity make GenAI outputs different from the information provided in books for liability purposes? 

\subsubsection{Counterfeit People}

The late philosopher Daniel Dennett thought one of the gravest harms from AI would stem from its ability to imitate humans without revealing it’s a human. “Our natural inclination to treat anything that seems to talk sensibly with us as a person—adopting what I have called the “intentional stance”—turns out to be easy to invoke and almost impossible to resist, even for experts.”\footnote{\href{https://www.theatlantic.com/technology/archive/2023/05/problem-counterfeit-people/674075/}{https://www.theatlantic.com/technology/archive/2023/05/problem-counterfeit-people/674075/}}  As Wired has recently reported, this is not a merely hypothetical scenario\footnote{\href{https://www.wired.com/story/bland-ai-chatbot-human/}{https://www.wired.com/story/bland-ai-chatbot-human/}}. The harm could be that it undermines the fabric of society: the ability to trust others. The entire economic system, the argument goes, is dependent on trust, as is democracy. Dennett compared counterfeit people to counterfeit money, which the government treats seriously because it, too, could undermine society\footnote{According to the government, over three-fourths of counterfeiters of all types serve a prison term with an average length of 16 months: \href{https://www.ussc.gov/sites/default/files/pdf/research-and-publications/quick-facts/Quick\_Facts\_Counterfeiting\_FY14.pdf}{https://www.ussc.gov/sites/default/files/pdf/research-and-publications/quick-facts/Quick\_Facts\_Counterfeiting\_FY14.pdf}}. If you could as easily be talking to a chatbot programmed by a con artist to impersonate a doctor, teacher, friend, or family member, it could throw sand in the gears of society, slowly impeding its progress, even if there is no single, concrete harm. Should tort law or some other law allow for such alleged harms because they may be too attenuated? Or should all AI be required by threat of severe penalties to conspicuously reveal it is not a human and to inject reminders with reasonable consistency?

\subsubsection{Safe Harbor}

On the other end, if a GenAI entity applies customary and best practices to prevent foreseeable harms to the extent such customs and best practices are known or knowable, should they receive a sort of safe harbor from most claims? Such a safe harbor may encourage wider adoption of preventative measures and instill a sense of industry norms, for example. 

\subsubsection{Data Laundering}

Suppose one Entity X is not allowed to access certain data for some reason (copyright, cease and desist order, paywall, etc.). Now suppose Entity Y can lawfully access and copy that same data, perhaps because of an exemption related to their tax status or nature of their work (e.g., academic research). Suppose further that Entity Y releases all that data as open source such that Entity X can simply download it from Entity Y. 

Is that a problem? If not, is it a problem if Entity X funds Entity Y’s data collection, knowing Entity Y will likely release it as open source? Does it matter if they’re partnered or in some sort of joint venture?

\subsection{Contract Law}

Contract law has had two prominent characteristics since the turn of the century: (i) the providers of the goods or services online are generally favored, and (ii) contract disputes are handled on a case-by-case, state-by-state basis. Neither of these characteristics may be ideal for GenAI. 

As to the first characteristic, most people will be familiar with clicking “I accept” whenever they sign up for a new website, service, update their phone apps, and more. Though it’s generally understood that nobody reads these terms, they are still deemed to be enforceable by courts in most instances. When a user must click a button saying they agree to terms before they are allowed to proceed, that it typically called “clickwrap,” which is different from when a website merely adds a hyperlink on the word “Legal” at the bottom of web pages that links to terms of service or terms of use, which is often called “browsewrap.” The former is an active assent, the latter is passive. Courts tend to favor the website so long as it can find that the visitor was or should have been aware of the terms, regardless of if the person actually looks at the terms and despite the terms often saying they are subject to change at any time on the whims of the provider, so it’s not even clear what the user is agreeing to. 

The passivity is what generally makes browsewrap unenforceable–but not always. This speaks to the second characteristic. It’s this gray area that requires clarification broadly applicable in all states. More pressing, GenAI’s particular reliance on web scraping from most of the public internet makes clarification around when a website’s terms of service is binding especially important. Do terms of service apply the same to bots as it does to humans visiting a site, regardless of how many times the bots visit (and therefore are likely “aware” of any terms, especially if they are visiting and scraping each page, which would include the terms of service page)? What if a bot is created to specifically avoid Legal pages so that the bot creator can claim plausible deniability of knowing about the terms? What are the minimum standards to make browsewrap applicable so it doesn’t have to be litigated on a case-by-case basis? 

In addition, can a site claim any rights just by putting them in a terms of service? For example, can it create property rights just by saying it’s so? If not, it may be helpful to clearly delineate common artifacts that property rights don’t attach to.

\subsubsection{Dataset Licensing}

Another issue revolves around licenses for datasets. A recent investigation into more than 1,800 datasets on popular dataset hosting sites determined that about 70 percent of the datasets “didn’t specify what license should be used or had been mislabeled with more-permissive guidelines than their creators intended.”\footnote{\href{https://www.washingtonpost.com/technology/2023/10/25/data-provenance/}{https://www.washingtonpost.com/technology/2023/10/25/data-provenance/}} 

This raises several questions, including the legality of unilaterally changing the license terms of a dataset, the enforceability of any illegality, whether intent should play a role, whether a party should be liable for not confirming the proper license is connected with the appropriate dataset, and how any license could reasonably be enforceable if the data can be modified and recombined with other data or into other formats. 

While loose adherence to documentation has become normalized in AI, and model and data cards are often sparse (perhaps increasingly so from large AI entities), it’s not clear that is the best approach from a market or policy perspective as it removes the ability to effectively police the use of datasets from people who wish to enforce the terms they are legally authorized to attach, and it absolves most parties of liability if they can honestly say they did not know better.

\subsection{Criminal Law}

\subsubsection{Intent}

Most criminal laws require both the physical act of committing the crime and the mental state of intending to commit the act. Notably, GenAI models do not have any intent when performing any actions. Developers create the GenAI without having any particularized knowledge of what the model will produce in response to a user prompt, users do not have any particularized knowledge of what the model will produce as a response to the user prompt, and the model itself has no desire or intent to produce anything in particular–it merely makes statistical associations between tokens to produce an output that has a high probability of being coherent. 

However, the user can use the output for any number of purposes, including criminal ones. 
The mental state requirement, therefore, might mean that while users could be found guilty of criminal conduct by using a model’s output, the developers and owners of the model (which could be different entities) may effectively be immune from virtually all criminal liability regardless of how harmful the act of the model. This would be the case despite models being given a task, assessing how to accomplish it, calculating the likely outcome, choosing which outcome to pursue, and then executing toward that goal, which starts to feel very much like “intent” as traditionally applied to humans. For example, GenAI may not only generate malicious code on request, but could also take steps to implement the code. In such circumstances should society consider the developers criminally liable, or should all risks be externalized to the users/society?

\subsubsection{Criminal Negligence}

Perhaps concepts such as criminal negligence could apply to model developers, which may not require intent by the developers or the model, but the standard would still require something like a negligent act that is so egregious by being foreseeably dangerous that it's likely to result in the risk of death or serious bodily harm. This is a high bar to meet because it can be impossible to know what is and is not foreseeably dangerous from the perspective of a GenAI model. Therefore, it could be extremely difficult (perhaps virtually impossible) for the government to prove to a jury beyond a reasonable doubt that the act was foreseeably dangerous.

\subsection{Property Law}

Property law comes in three large buckets: real property (like land), personal property (other physical objects you own, like jewelry), and intellectual property (intangible property like patents and copyrights). It’s unclear if any personal property rights do or should attach to intangible property on the internet. 

For example, if you own a physical book consisting of a story you wrote and someone steals it, they are violating your right of possession, right of exclusion, right of enjoyment, right of control, and right of disposition. They are not violating your copyright. 

Conversely, if there is a digital version of your book online and someone steals it, they are violating your exclusive right to reproduce the work under copyright (because they necessarily had to make a copy of the file), but they may not be violating any personal property rights because it’s unclear whether any of them apply to digital property. In fact, in most cases, any data online that people do not take steps to keep secret, private, or exclusive may have no protections other than copyright. 

Because GenAI developers must make copies of every file it finds on the web that they want to include in their training dataset, they are likely committing copyright infringement (though it may not be illegal if it’s considered fair use, as discussed above). However, if personal property rights extend to digital property, they may also be committing theft, larceny, conversion, or a number of other property-related acts that could trigger civil or criminal liability. 

\subsubsection{Robots.txt}

If there is a right to exclude, that could also trigger claims based on robots.txt, which is a protocol that tells bots which web pages it’s allowed to visit on a given site. Currently, adhering to robots.txt is voluntary, but if digital content is property with a right to exclude, then claims such as trespassing could apply when a developer ignores robots.txt and scrapes data from additional web pages despite the webmaster expressly asking them not to. Claims could become even stronger if a company that says it respects the Robots Exclusion Protocol but doesn’t follow it, because it could lead to consumer protection, unfair advertising, or deceptive trade practices claims. Without property rights (either personal or intellectual), though, such claims may be challenging to sustain.

Things become trickier when considering enforcement. Even if, for example, a site blocked for-profit bots from scraping its content, the developers of those bots could probably obtain the content by just using the publicly-available datasets created by nonprofit research organizations who scraped the same websites. Should for-profit entities be responsible for ensuring they don’t acquire the data through any means, or only by not scraping the original site itself?

\subsubsection{Publicly Available Data}

Publicly available isn’t the same as public domain. This may be especially relevant for the content on the public internet that was not intended for that space. For example it may have been stolen from elsewhere via hacking and then placed on the public internet. Or, the information may have been shared in confidence, intended for a small audience such as a group chat, but then someone from the group posts it on the public internet. Or it could be from an angry employee who posts trade secrets in a public space. 

The fact that some data is publicly available does not, by itself, tell us anything about whether it was ever intended to be publicly available. Should that matter?

\subsubsection{Opting Out}

Aside from privacy laws, suppose a GenAI entity has an opt-out form. When a property owner requests that a GenAI entity remove the owner’s data from a training set or the GenAI model, must the entity honor the request? Must the entity even offer opt-out, or is it just a nice-to-have or perhaps a mitigating factor in relation to other claims? Should opt-out requests apply retroactively, regardless of how old the dataset or model is?  Or should the more stringent requirement of opt-in be required to use someone’s data?

\subsection{The First Amendment}

The First Amendment protects the freedom of speech and expression, placing a high legal bar on government actions that limit such speech. The two standards courts apply when reviewing restrictions on speech are intermediate scrutiny (where the law must further an important government interest and must do so by means that are substantially related to that interest) and 
strict scrutiny (where the law must further a "compelling governmental interest," and it must be narrowly to achieve that interest). 

If GenAI outputs are the speech of the developers, or the model itself, then regulating GenAI will require overcoming the scrutiny standards, which could be difficult or impossible, depending on the regulation. This may be the first time there is serious debate about whether a non-human’s outputs are constitutionally protected speech, and the outcome could be significant. For example, a law that prevents GenAI from disparaging ethnic groups could violate protected speech, making the law unconstitutional. However, it may be that GenAI outputs are not speech at all, which means there is no speaker, so there is no receiver of speech, and therefore the government would have a wide latitude to regulate GenAI as it sees fit. 

\subsubsection{CDA 230}

Section 230 of the Communications Decency Act states, in relevant part, that “No provider or user of an interactive computer service shall be treated as the publisher or speaker of any information provided by another information content provider.”\footnote{\href{https://www.law.cornell.edu/uscode/text/47/230}{https://www.law.cornell.edu/uscode/text/47/230}} For all prior platforms and types of AI, companies could usually claim the company itself (the “provider” of the “interactive computer service”) was not the speaker or publisher. Rather, they were merely hosting content provided by others. This is the case with Facebook, YouTube, and other social media platforms, for example. 

But now that GenAI allows companies to create entirely new sentences, it may be that sites like Perplexity.AI and ChatGPT, and features like Google’s AI Overview, are the speakers, which could remove the liability shield of CDA 230 and make the companies themselves liable for what the GenAI posts.  



\section{Conclusion}

The rapid evolution and adoption of GenAI presents a unique challenge to existing legal frameworks. As GenAI systems diverge from traditional software in their scale, scope, and capacity for both beneficial and harmful applications, existing laws, many drafted in the pre-digital era, struggle to address the novel issues presented. We examined key areas where legal doctrine may require significant revision or reinterpretation, including copyright, privacy, torts, contracts, criminal law, property law, and the First Amendment.

Rather than advocating for specific legal changes, we highlight critical questions policymakers must confront as GenAI technology continues to advance. Central to this discussion is a need to balance the potential benefits of GenAI with the urgency of mitigating potential harms before it’s too late, which has led to unfortunate outcomes by other technologies like autonomous vehicles and social media\footnote{\href{https://www.nytimes.com/2018/10/15/technology/myanmar-facebook-genocide.html}{https://www.nytimes.com/2018/10/15/technology/myanmar-facebook-genocide.html}}\footnote{ \href{https://www.lemonde.fr/en/le-monde-africa/article/2022/08/18/burkina-faso-condemns-ethnic-cleansing-messages-circulating-on-whatsapp\_5994006\_124.html}{https://www.lemonde.fr/en/le-monde-africa/article/2022/08/18/burkina-faso-condemns-ethnic-cleansing-messages-circulating-on-whatsapp\_5994006\_124.html}}\footnote{ \href{https://www.washingtonpost.com/politics/2020/02/21/how-misinformation-whatsapp-led-deathly-mob-lynching-india/}{https://www.washingtonpost.com/politics/2020/02/21/how-misinformation-whatsapp-led-deathly-mob-lynching-india/}}. As GenAI becomes increasingly integrated into various aspects of society, proactive engagement with these complex legal and ethical questions is paramount.

\section*{Acknowledgements}
Special thanks to Crystal Nam and Chandler Lawn for their useful feedback.

\bibliography{references}
\bibliographystyle{abbrvnat}

\end{document}